# Observation of d-wave scaling relations in the mixed-state specific heat of $YBa_2Cu_3O_7$.

B. Revaz, J.-Y. Genoud, A. Junod, A. Erb and E. Walker

Département de Physique de la Matière Condensée, Université de Genève

24 quai Ernest-Ansermet, CH-1211 Genève 4 (Switzerland).

## Abstract

The low temperature specific heat of a high purity $YBa_2Cu_3O_{7.00}$ single crystal grown in a $BaZrO_3$ crucible is measured from 1.2 to 10 K in magnetic fields from 0 to 14 T. The anisotropic component of the excess specific heat due to field, $C_{aniso}(T,B) \equiv C(T, B//c) - C(T, B \perp c)$, is obtained directly from experiment. $C_{aniso}(B)$, which is free from background due to phonons, paramagnetic centres, residual normal-state regions, etc., is a characteristics of the vortex system. $C_{aniso}(B)$ is shown to follow a scaling relation predicted recently for line nodes characteristic of d-wave vortices. Our experimental field and temperature range corresponds to a crossover region for the scaling function where neither the low-$T/B^{1/2}$ limit, $C_{aniso}(B) \propto TB^{1/2}$, nor the large-$T/B^{1/2}$ limit, $C_{aniso}(B) \propto B$, strictly apply. The variation of the entropy caused by the magnetic field at low $T$ is compared with measurements near $T_c$. Thermodynamic compatibility is found at the qualitative level.

Whereas many experiments sensitive to the surface of the sample tend to establish that the symmetry of the order parameter is d-wave in high temperature superconductors (HTS) [1], it is felt that a general consensus should be also supported by bulk measurements. Specific heat ($C$) experiments are insensitive to the phase of the order parameter, but they can provide bulk information on the behaviour of the density of states (DOS, $N(E)$) near the Fermi level $E_F$. The electronic part of $C/T$ is proportional to $N(E)$ averaged over an interval $\approx k_B T$ around $E_F$. For a d-wave symmetry, the gap vanishes on lines of nodes on the Fermi surface, therefore the DOS averaged over all directions in the reciprocal space behaves as $N(E) \propto |E|$ (with the convention $E_F \equiv 0$). This implies $C_{electron}/T \propto T$ at $T \ll T_c$ in zero magnetic field, whereas for a fully gapped superconductor $C_{electron}/T \approx 0$ in the same conditions.

This simple criterion can hardly be applied in practice, because the low temperature specific heat of HTS in zero field contains many contributions that have an ill-defined variation with $T$. Examples are extrinsic magnetic contributions due to small numbers of weakly interacting paramagnetic centres, causing low-$T$ « upturns » [2], and the so-called « linear term », which was reported to have a more complicated $T$-dependency with a fractional exponent [3] [4]. The lattice contribution itself is not fully identified, as suggested by the widely dispersed values of the Debye temperature reported in the literature [5]. Therefore the detection of a minority contribution $C_{electron}/T \propto T$ has to be based on indirect techniques such as differential measurements versus doping [6] [7], whereas multiparameter fits of the total specific heat give more ambiguous results [8] [9] [10]. Note also that the result $C_{electron}/T \propto T$ is not general for d-wave symmetry, and may be modified by the angular dependence of the gap in the reciprocal space and by s+d mixing.

Another route consists in studying the variation of the specific heat versus the magnetic field $B$. According to a simplified argument, the energy of carriers circulating around a vortex is shifted by Doppler effect. This shift has a dramatic effect on the DOS when the gap is small, so that the essential contribution comes from the vicinity of the nodes. In a low temperature limit that will be specified later, the DOS at $E_F$ becomes proportional to $B^{-1/2}$ for one vortex, therefore $N(E_F) \propto B^{1/2}$ for the full vortex system [11]. It follows that $C_{vortex}/T \propto B^{1/2}$. This term is still a small fraction of the total specific heat, except in the vicinity of 1 K at high fields. Although the lattice contribution does not interfere here because it does not depend on $B$, various species of paramagnetic centres add their own variation with $B$ [2] [12]. Multiparameter fits were used by Moler *et al*. [8], Fisher *et al*. [9] and Wright *et al*. [10] to pick out the contribution of interest. Both groups concluded in the presence of a d-wave term. Moler *et al*. based their analysis on $C(T,B)$ data for $2 \leq T \leq 7$ K and $0 < B \leq 8$ T, Fisher *et al*. and Wright *et al*. on data for $0.5 \leq T \leq 10$ K and $0 < B \leq 9$ T.

In order to get rid of the uncertainties introduced by multiparameter fits, essentially due to the fact that the analytic form of some contributions is unclear, we look at the *anisotropic* component of the specific heat in a magnetic field $C_{aniso}(T,B) \equiv C(T, B//c) - C(T, B \perp c)$. We measure the low temperature specific heat with $B//c$, then rotate the sample and measure it again with $B \perp c$. The sample holder and thermometer remain in the same position with respect to the field, so that calibration errors tend to cancel out. The lattice specific heat, the zero-field d-wave term $C_{electron}/T \propto T$, the nuclear specific heat, other paramagnetic contributions, and possible residual normal-state contributions also cancel out. $C_{aniso}(T,B)$ is a fraction of the vortex specific heat for $B//c$ that would approach 100% in the limit of large anisotropy. The anisotropy factor of the upper critical field is $\Gamma \equiv (m_c/m_{ab})^{1/2} = 5.3 \pm 0.5$ for the present overdoped crystal [13], so that we still obtain a large fraction of the full vortex specific heat. An advantage is that the latter is directly measured, rather than modelled in a fit, so that the experiment gives not only the amplitude, but also the shape of the vortex specific heat at low temperature. We show that its $B$ and $T$ dependence obeys the scaling relation recently given by Volovik [14] and by Simon and Lee [15] for d-wave vortices. In contrast to earlier findings [8] [9] [10], we find that the usual experimental field and temperature range corresponds to a crossover region where neither the low $T$, high $B$ limit $C_{vortex}(B) \propto TB^{1/2}$, nor the opposite limit $C_{vortex}(B) \propto B$ strictly apply. The scaling property nevertheless shows that bulk thermodynamic properties support the d-wave scenario, inasmuch as methods which are sensitive only to the amplitude of the order parameter can.

Recently, YBa$_2$Cu$_3$O$_7$ single crystals could be grown in inert BaZrO$_3$ crucibles [16]. The new level of purity of such samples, 99.995% [17], has allowed successful experiments such as imaging the vortex lattice by scanning tunnelling microscopy [18], suppression of the magnetic « fishtail » effect [19], observation of specific heat peaks at the flux line lattice melting transition in fields up to 16 T [13], etc. The 18 mg single crystal used for the present study (and also in Refs. [13] and [19], code AE195G) was fully oxygenated in 100 bars O$_2$ at 300°C. The calorimetric transition midpoint $T_c \cong 87.5$ K [13] was typical of overdoped YBa$_2$Cu$_3$O$_{7.00}$. The high chemical purity and the low concentration of oxygen vacancies should both favour the observation of d-wave properties in the clean limit ; such requirements were also found to be essential for the calorimetric observation of a transition in the vortex lattice [13]. Overdoped crystals from the same breed have shown a quasi-linear variation of the condensate density with $T$ over a wide $T$ range, which is not the case at optimal doping [20].

The low temperature specific heat was measured for $1.2 \leq T \leq 10$ K and $0 \leq B \leq 14$ T, using a thermal relaxation technique [12] [21]. The absolute accuracy is estimated to 3%. The present differential experiments rely on reproducibility, which is one order of magnitude better. Figure 1 shows the raw data in $B = 8$ T. The full lines give the total heat capacity, including the sample holder, thermometer, silicone grease and crystal, for $B//c$ and $B \perp c$. The quantity of interest is the difference, which becomes more and more significant at lower temperature. The relevant uncertainty margin is illustrated by two independent runs in zero field, one belonging to the series $B//c$, and the other to the series $B \perp c$ after rotation. Ideally their difference should be zero. Practically it remains small with respect to the differences that we want to study in $B > 0$ (Fig. 2).

The low temperature specific heat in $B = 0$ is shown in the usual $C/T$ versus $T^2$ plot in the inset of Fig. 1. There is no upturn down to 1.2 K. Independent measurements in a dilution refrigerator [22] confirmed the absence of any upturn down to 0.2 K, a temperature below which the hyperfine nuclear contribution takes over. This exceptional quality facilitates the analysis of intrinsic terms. The field, temperature and doping dependencies of $C(T, B \perp c)$ give reasons to think that the *apparent* residual term linear in $T$ results from a superposition of terms with a more complicated $T$-dependence, including a d-wave component. This point will be discussed in a forthcoming publication [23].

Strictly speaking, the anisotropic component of the specific heat may contain a spurious contribution from residual paramagnetic centres if the Landé factor $g$ is anisotropic. We proceed to show that this may be safely neglected, because the extrinsic $S = 1/2$ Schottky anomaly $C_{Sch} = Nx^2 e^x / (e^x + 1)^2$, $x \equiv 2gS\mu_B B / k_B T$, presumably due to copper ions sitting next to an oxygen vacancy and/or to paramagnetic impurities, is small for the present pure and overdoped crystal. The amplitude $N$ was estimated using $C(T, B = 4\,\text{T}) - C(T, B = 14\,\text{T})$, $B \perp c$. The Schottky peak occurs in a temperature region where it is best identified in a $\approx 4$ T field. The reason for choosing the maximum field, 14 T, rather than the $B = 0$ curve as a reference, is that the latter is complicated by magnetic interactions, whereas the former is smooth (using alternatively the $B = 0$ curve, we would optimistically conclude that $N$ is unmeasurably small). Finally the vortex contribution for $B \perp c$ is small. In these conditions, we conservatively estimate $N = 0.65$ mJ/gat K, i.e. 0.03% of spin ½ per Cu atom. A correlation with the concentration of oxygen vacancies appears when this result is compared with the values (in the same units) $N = 1.8$ for the YBCO$_{6.95}$ crystal of Ref. [8], and $N = 2.5$ for the YBCO$_{6.92}$ crystal of Ref. [24], also grown in BaZrO$_3$. A similar correlation was observed for high purity ceramics [12]. Using the measured $N$ and the anisotropic Landé factor as given by electron paramagnetic resonance [25], we calculate the anisotropic component of the Schottky contribution in 8 T (Fig. 2, full line). It represents less than 5% of $C_{aniso}$ in the same field.

Therefore $C_{aniso}(T, B)/T$ shown in Fig. 2 for $B = 0$, 1, 2, 4, 8, 10 and 14 T represents essentially the vortex specific heat $C_{vortex}(T, B//c)/T$, within a numerical factor $f(\Gamma)$ that will be precised later. Two distinct features appear in these direct measurements. First, the vortex specific heat is not truly linear in $T$, in which case the data would align on horizontal lines. $C_{aniso}/T$ decreases with $T$, and this effect is enhanced by the field. Second, the specific heat does not increase linearly with $B$, as would be the case for the vortex core contribution of low-$T_c$ type-II superconductors [26], but increases with a smaller exponent. This is particularly evident at the low temperature end, *e.g.* for the 4 and 14 T curves.

The variation of $C_{aniso}/T$ with $T$ is in fact necessary for thermodynamic consistency at low and high $T$. The entropy is field-independent both at $T=0$ and $T>T_c$, so that its anisotropic component $S(T,B//c)-S(T,B\perp c)$ can be measured at any intermediate temperature by integration of $C_{aniso}/T$ starting either from $T=0$ or from $T>T_c$. The curves have to match. As shown in Fig. 3, this would not be possible if $C_{aniso}/T$ kept its low-temperature value.

The additional free energy due to d-wave vortices was calculated by Volovik [11], Kopnin and Volovik [27], and Volovik [14]. Two regimes are described, depending on the variable $x \equiv (T/T_c)/(B/B_{c2})^{1/2}$; $B_{c2}$ is the upper critical field. In the low temperature limit $x \ll 1$, the vortex specific heat is linear in $T$ with $C_{vortex}/k\gamma_n T_c = (T/T_c)(B/B_{c2})^{1/2}$, $\gamma_n$ the Sommerfeld constant and $k$ a numerical constant of order one ($k = 1.605$ according to Ref. [28]). In the small field limit $x \gg 1$, $C_{vortex}/k\gamma_n T_c \propto B/B_{c2}$. The anisotropy enters through $B_{c2}$. With $\Gamma = 5.3$, the quantity $C_{aniso}$ that we measure represents $f(\Gamma) = 1 - \Gamma^{-1/2} \cong 57\%$ of $C_{vortex}$ in the small-$x$ regime, and $f(\Gamma) = 1 - \Gamma^{-1} \cong 81\%$ in the large-$x$ regime. In the following, we assume that $f(\Gamma)$ depends only weakly on $x$.

At variance with the assumption made earlier [8] [9] [10], the data for $1.2 < T < 10$ K are not in the $x \ll 1$ regime because $C_{aniso}/T$ varies with $T$ even in the highest fields (Fig. 2). It is equally evident that they are not in the $x \gg 1$ regime where linearity in $B$ is predicted. This makes bulk confirmation of d-wave properties uncertain at first view. However, it was shown recently by Simon and Lee [15] and by Volovik [14] that the d-wave vortex specific heat should obey a scaling relation $C_{vortex}/TB^{1/2} \propto F_c(x)$, independently of the regime. The asymptotic limits of the scaling function are $\lim_{x \ll 1} F_c(x) = 1$ and $\lim_{x \gg 1} F_c(x) = 1/x$. An empirical interpolation function $F_c(x) = (1+x)^{-1}$ was proposed [14]. This scaling should remain valid as long as the gap is constant, i.e. for temperatures not too close to $T_c$.

The d-wave scaling relation $C_{vortex}/TB^{1/2} \propto F_c(x)$ is tested in Fig. 4a. An equivalent scaling form, $C_{vortex}/T^2 \propto F_c(x)/x \equiv F_c^*(x)$, is less critical and possibly overly suggestive (Fig. 4b). All data previously shown in Fig. 2 tend to cluster onto a single curve, which may be *empirically* described by a logarithmic law $F_c(x) = a - b\ln x$. The scatter increases at the high temperature end because of the large phonon background. The interpolation scaling function $F_c(x) = (1+x)^{-1}$ [14] is shown in Fig. 4a with the fitted parameters $B_{c2} = 320$ T and $f(\Gamma)k\gamma_n = 1.5$ mJ/gat K$^2$. These parameters are correlated. Somewhat different combinations with the same value of the ratio $f(\Gamma)k\gamma_n/B_{c2}^{1/2}$ are acceptable. Literature values $k = 1.605$ [28], $\gamma_n \cong 1.5$ mJ/gat K$^2$ [29], $B_{c2} \cong 140$ T [30], together with $f(\Gamma) = 0.57$, overestimate this ratio by about one third, but $\gamma_n$ and $B_{c2}$ are evaluated by indirect methods at higher temperatures and are therefore subject to uncertainty. The result of Moler *et al.* [8], $C_{vortex}/TB^{1/2} = 0.91$ mJ/mol K$^2$ T$^{1/2}$, is shown as a horizontal line in Fig. 4a, taking into account the relevant factor $f(\Gamma) \cong 0.62$ estimated for their YBa$_2$Cu$_3$O$_{6.95}$ crystal. It represents an average behaviour, as can be expected from a technique based on a fit in the 2-7 K region, for a crystal that has a presumably smaller DOS owing to its lower oxygen content [29]. According to Fig. 4, the true asymptotic behaviour $C_{vortex}/TB^{1/2} = const$ can be reached only below $T/B^{1/2} \approx 0.1$ K/T$^{1/2}$, e.g. at $T < 0.3$ K in $B = 8$ T and still lower temperatures in lower fields. This is a domain where the nuclear background becomes large. Therefore the direct observation of a $TB^{1/2}$ term appears to be more problematic than it was expected.

The entropy diagram shown in Fig. 3 for $B = 8$ T imposes some bounds on the anisotropic specific heat at higher temperature. Starting from $T > T_c$, the entropy cannot be measured very far below $T_c$ with the present mg-size crystal, because small errors in the absolute value tend to wind up on a long integration path. However, these high-temperature data determine the *average* coefficient of the linear term between 0 and 60 K, $\langle C_{aniso}(T, B = 8\,\text{T})/T \rangle \cong 0.15$ mJ/gat K$^2$. This is significantly smaller than the low temperature limit $C_{aniso}(0, B = 8\,\text{T})/T \cong 0.24$ mJ/gat K$^2$, which gives rise to the entropy represented in Fig. 3 by a straight line. Alternatively, the scaling function $F_c(x) = (1+x)^{-1}$ corresponds to a d-wave entropy $\propto \ln(1+x)$, shown by the lower curve of Fig. 3. A quantitative extrapolation of this scaling function much above 10 K is uncertain, but it appears at least to leave correctly some room for the thermally excited vibrations of the vortices and for the entropy « peak » associated with the vanishing gap near $T_c$. Therefore the decrease of the ratio $C_{vortex}/TB^{1/2}$ as a function of the temperature, which is predicted by theory and now observed (Figs. 2 and 4a), is necessary for thermodynamic consistency.

The collapse of the specific heat data on a single scaling function without any background subtraction gives bulk, objective support to the existence of line nodes in the gap function of YBa$_2$Cu$_3$O$_{7.00}$. Scaling in the variable $x \propto T/B^{1/2}$ is quite exotic a signature, that differs markedly from the usual scaling of spin or band magnetism in the variable $T/B$; it is a characteristics of nodes. The *amplitude* of $C_{vortex}/T$ generally scales with $B^{1-D/2}$, where $D$ is the dimension of the nodes [14]. Because here $C_{vortex}/T \propto B^{1/2} F_c(x)$, the collapse of the data in Fig. 4 further indicates that the dimension of these nodes is one, i.e. they are lines.

Analogous measurements for Bi$_2$Sr$_2$CaCu$_2$O$_8$ are in progress. The vortex specific heat is found to be very similar to the present YBa$_2$Cu$_3$O$_7$ data, both from the point of view of the amplitude (per unit volume) and its variation with the field [23], in spite of the difference in the rigidity of the vortices. A field on the order of 0.1 T decouples the vortices of Bi$_2$Sr$_2$CaCu$_2$O$_8$ into nearly independent pancakes [31], in contrast to YBa$_2$Cu$_3$O$_7$ where the decoupling field is rather on the order of 10 T [32]. Therefore thermally excited vibrations of vortices are expected to yield very different contributions to the specific heat in these two systems. The fact that the low-$T$ mixed-state specific heat is similar tends to rule out such contributions, and to reinforce the conclusion that the mixed-state specific heat is indeed governed by a finite DOS at $E_F$ around the vortices.

*Acknowledgements* - The authors are grateful to G.E. Volovik, S. Simon, L. Pintschovius, W. Reichardt, J.M. Triscone, L. Fabrega, J. Muller and J. Sierro for fruitful discussions, to K. Neumaier and S. Sridhar for providing data prior to publication, and to A.

Naula for technical assistance. This work was supported by the Fonds National Suisse de la Recherche Scientifique.

## Figure captions

Figure 1. Total heat capacity $c/T$ (including the sample holder) in a field $B = 8$ T. Upper curve : $B//c$. Lower curve : $B \perp c$. Inset : specific heat of the sample $C/T$ versus $T^2$ in $B = 0$. One gat = 1/13 mol = 51.25 g.

Figure 2. Anisotropic specific heat $C(T, B//c)/T - C(T, B \perp c)/T$ versus $T$ (note the logarithmic scale), from top to bottom 14, 10, 8, 4, 2, 1 and 0 T. Full line : calculated Schottky anomaly in $B = 8$ T with $g_{//} = 2.208$, $g_{\perp} = 2.047$, $S = 1/2$, $N = 0.65$ mJ/gat K (see text).

Figure 3. Anisotropic component of the entropy $S(T, B//c) - S(T, B \perp c)$ in $B = 8$ T. Straight line : integration of $C_{aniso}/T$ upwards with $F_c(x) = 1$. Curved line : $F_c(x) = (1+x)^{-1}$. In both cases the parameters are taken from the fit of Fig. 4a. Open circles : integration of $C_{aniso}/T$ from 95 K downwards, using high temperature data for the same sample [13]. The entropy peak near $T_c$ in the overdoped state is about twice as high as for the optimally doped state [12] ; this is confirmed by a measurement with a second high-purity crystal [33].

Figure 4. (a), scaling plot of the anisotropic specific heat $C_{aniso}/TB^{1/2}$, same data as for Fig. 2 ($1.2 \leq T \leq 10$ K, $1 \leq B \leq 14$ T). Full line, scaling function $F_c(x) = (1+x)^{-1}$ with $B_{c2c} = 320$ T and $f(\Gamma)k\gamma_n = 1.5$ mJ/gat K$^2$. Dashed line, function used in Ref. [8], $C(T, B \perp c)/TB^{1/2} = 0.91$ mJ/mol K$^2$ T$^{1/2}$. In the latter case, we have assumed $f(\Gamma) = 0.62$, i.e. $\Gamma = 7$ for YBa$_2$Cu$_3$O$_{6.95}$. (b), scaling plot of the anisotropic specific heat $\Delta C_{aniso}/T^2$, showing the scaling function $F_c^*(x) \equiv F_c(x)/x$.

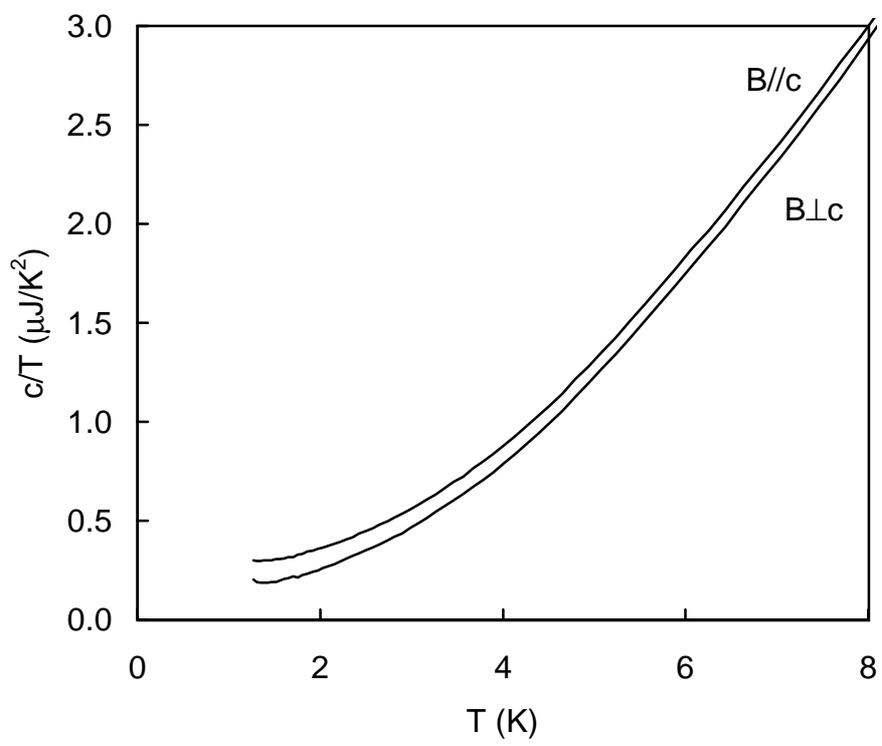

figure 1

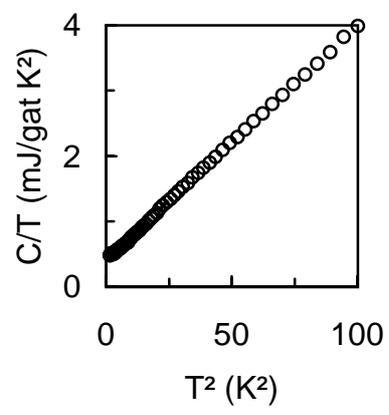

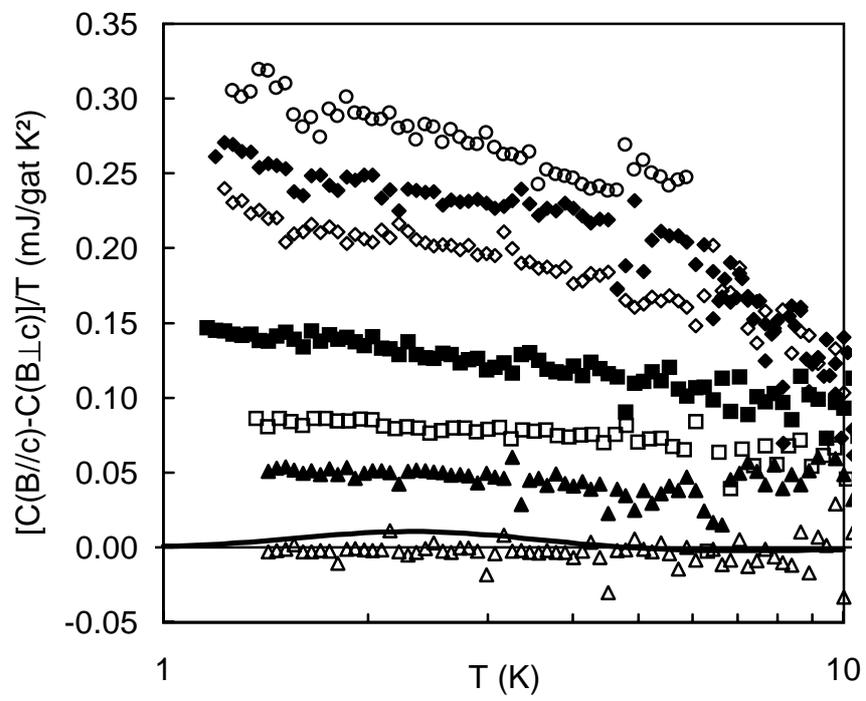

figure 2

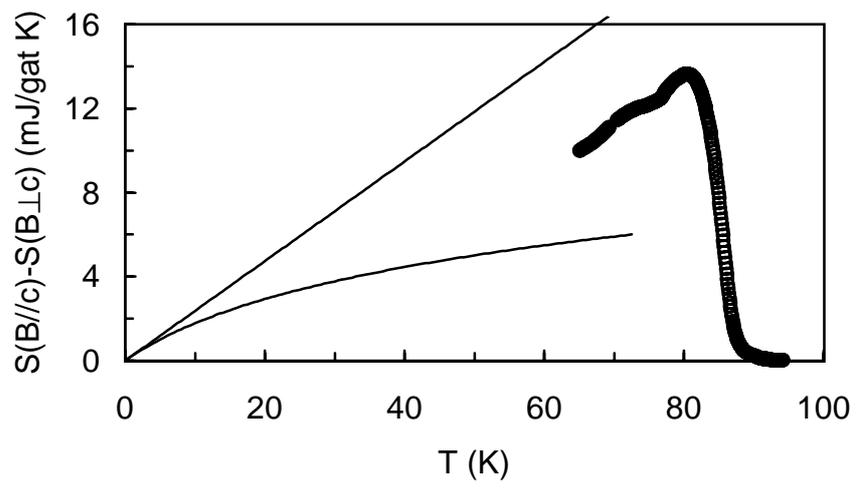

figure 3

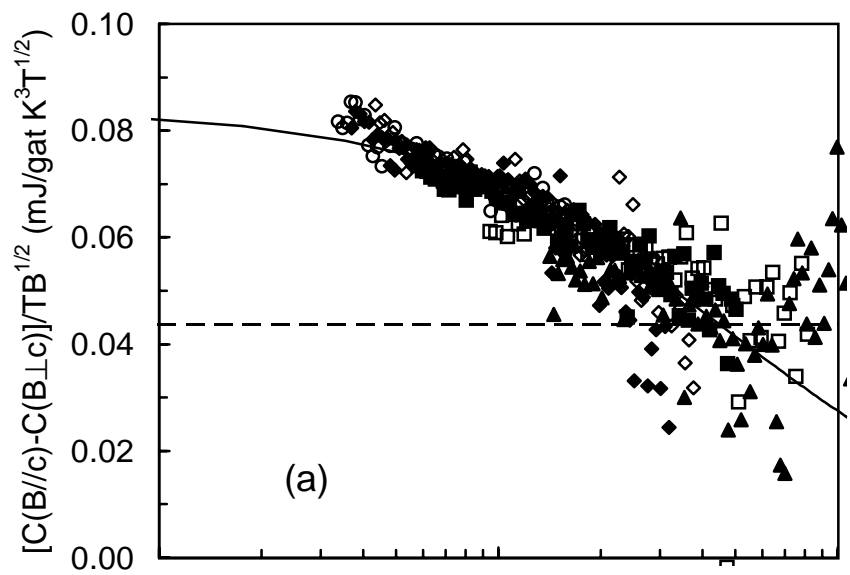

figure 4a

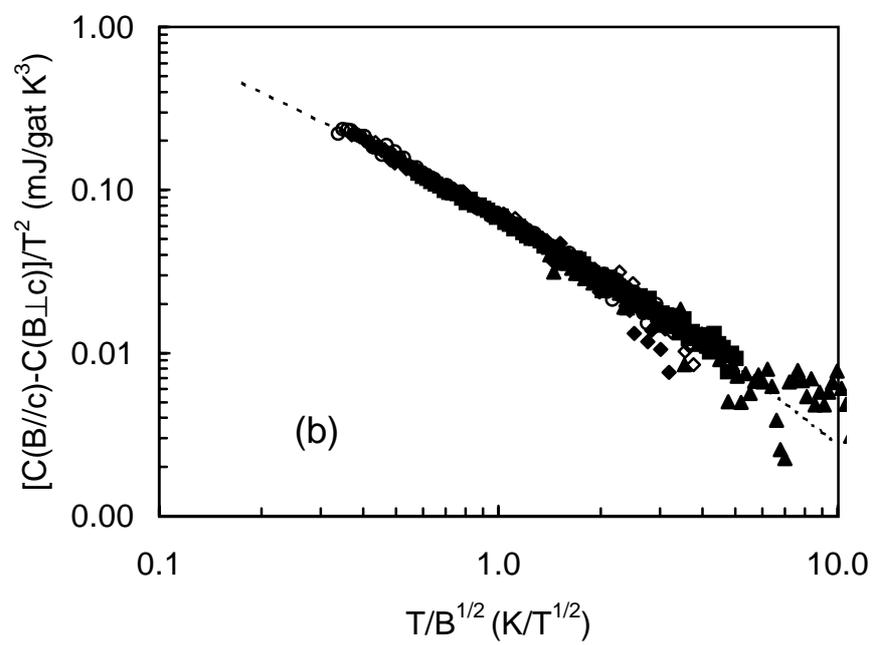

figure 4b